# SCHEDULING INTELLIGENT SYSTEM FOR TIME SHORTENING

**Gabriela Prostean, Octavian Prostean, Iosif Szeidert, Ioan Filip**
**"Politehnica" University of Timisoara, Romania**
gprostean@eng.upt.ro, iosif.szeidert@aut.upt.ro

## ABSTRACT

*The paper presents a scheduling intelligent system intended for the project management and for the operation management as well, having integrated a planner time buffer method combined with the PERT (Programme Evaluation and Review Technique) method which can drastically short the planned time. The system also adjusts if necessary the duration for the un-expecting situations during the evolution of the planner recalculating the probability to reach the deadline. The system is developed with a friendly graphical interface, which guide the user during the progress of the project providing warnings and suggestions for adjusting in real time the planner. Once the scheduling intelligent system is launched in progress, its functions are combined at the different levels, depending of the user needs. The base functions of the system are: planning, diagnosis, supervising and forecast. A real implementation is showed as a study case, is related to a software development planner.*

## INTRODUCTION

In case of leading technologies industries, the products life cycle no longer observes the classical curve profile, it changing into a « saw tooth » profile. Before the close of the stage of a leading technology product throwing on the market, that product is already considered morally obsolete because a new, more competitive product is coming forth. The rhythm of the new types of products launching on the market is so quick that the producing firms are permanently faced with the danger of losing the already gained position on the market. This new background led to the study of some more efficient techniques for the Management of the Projects for new products development, which were meant to represent a decision support in the management of the products life cycle.

An Intelligent System conceiving and development in such a field as that of the development Projects Management of the new products require complex knowledge and experience in many fields, such as: design, production, programming, computers, and the economic experience of the commercial companies. The capacity to make a clear distinction between the tasks, the knowledge of the respective field, the problems solving method, as well as the integration of a well structured conceptual model, stand for the basis required for the development of such an Intelligent System, by taking into account the uniqueness of each type of planner.  The scheduling intelligent system called "PManager"  and presented in this article comprises a new method of present interest in scheduling called "Time Buffer" conceived for shortening the time required by the project achievement on the basis of the time reserve assigned to the Critical Path, which is calculated according to the following rule: there is calculated the sum of the safety estimations obtained through the differences between the pessimistic and optimistic periods (estimated according to PERT method) of the critical tasks, and then it is reduced to a half [6], the technique of time buffer obtaining being inspired from the Theory of Constraints developed by Elyahu M. Goldratt [1].

PManager Intelligent System enables carrying out the probabilistic evaluation of integration in the planned time, according to PERT method, the synchronization of the project evolution, the supervising of the tasks according to the optimistic periods (the strictly operational ones), and the adjustment of the tasks evolution in real time by means of the "Time Buffer". The aim of the development of this Intelligent System in planning is to integrate within the structure of such a program product a new mechanism for the drastically shortening of the period for the projects carrying out, offering real time reference, adjustment and control possibilities.[9]

The system has been achieved by means of MS Visual Studio C++ medium. The graph of the evolution of the planned and real costs for the project is generated. The adjustments carried out for each type of planner are stored in the knowledge base as frames, and subsequently used for the realization of the same type planners, for the warning of the user concerning future unexpected situations.

## INTELLIGENT SYSTEMS MODULES

- "Project management engineering" module

The feasibility of the project is analyzed within this module, the managers being thus offered the required support in conceiving the plan of tasks. The module practically stands for the system "brain". There are used the information previously processed by means of Microsoft Project software, and the knowledge available in the system knowledge base. The module offers the user the possibility of selecting the knowledge base corresponding to the project nature, project type, time constraints, etc. When there are no data entered by the user, default values can be assigned. [3][4]

The study of feasibility is activated by using the relevant information and knowledge from the knowledge base. In conjunction with the project feasibility study, the system generates a plan which details the planning of the tasks, followed by financial resources assignment. [2][10] By means of system – user interaction, a list of tasks planned according to PERT method is recursively built, the probability factor for the project carrying out being calculated along with the data entering. [6]

- "Progress supervising" module

The project monitoring phases compare the planner current status with the planning estimations and issue warnings when there appear nonconformities. In case there appears a deviation from the optimistic period of PERT planner, the user can adjust that period from the project total time reserve (the time buffer), calculated on the basis of the following rule: there is calculated the sum of the safety estimations obtained through the differences between the pessimistic and optimistic periods of the critical tasks, and then it is divided to two,[5][6]. Following the adjustment of the period of an task, the remaining project will be replanned in real time. The supervising and adjustment of the planner evolution in real time can be achieved only on the basis of a synchronization of the periods evolution with the system clock. That is why the user shall communicate the following synchronization data to PManager Intelligent System, (Figure 2):

● The date of the project beginning
● The work program
● The hour of the project starting
● The time unit chosen for the tasks planning (minutes, hours, days, months)

The weekend days are not considered work days, and as for other possible free days (holidays, vacation, etc.), the planner evolution can be stopped.

Thus, a project shall be replanned in case the existing plan proves to be inadequate or not valid. All the assumptions used for the plans generation can be checked by the user within any stage, along the plan generation or project monitoring.

- "Updating" module

At the end of each project, the knowledge from "Experience" module are examined and used for updating the system knowledge base. This can be achieved by the active participation of the user by means of "Updating" module. In fact, the choice belongs to the user, whether to ignore a certain piece of knowledge, as well as an exception condition, or to include it in the knowledge base.

- "Experience" module

"Experience" module grants PManager Intelligent System a greater capacity of anticipation than that of a human specialist.

It is possible that the planner may not be able to develop according to the initial network; a case when the manager can chose to enter some additional periods for specific situations, regenerating a new type of plan. These modifications will be automatically entered in 'Experience' module. The knowledge obtained through 'Experience' module are used in feasibility studies, and in the tasks planning in a more efficient way for subsequent projects. In case of a normal course of events, the system can supply a current display, or it can print current reports during the various stages along the project progress, or it can highlight what was left to be executed.

- "Time buffer" module

This method ensures the implementation of "time buffer" method – proposed in this paper. This method supposes taking over only the optimistic periods of the planner from PERT method, according to which the project will be monitored, the critical path being calculated. A (safety) time reserve for the sequence of the optimistic planner critical tasks will also be calculated on the basis of the data implemented in PERT method. This time reserve represents the time buffer, (the calculation rule for the time buffer is presented in "Progress supervising" module). The system further offers the possibility of supervising the planner's evolution in real time, its follow up being according to the optimistic periods (periods considered to be strictly operational). Thus, a drastically shortening of the period of the whole planner, and costs minimization, implicitly, are aimed at, an operational method in scheduling being thus conceived.

Obviously, inherent, unexpected situations can appear during the evolution of the planner. In such cases, the system offers the possibility of transferring an additional period from the time buffer for actual delayed tasks. The planner, the network, respectively, is automatically updated, the critical path being recalculated, and the changes together with the causes that generated them will be memorized in the system knowledge base by means of "Experience" module.

The system can further offer a complete diagnostic of the evolution of each task of the planner, a model being thus created on the basis of which estimations as close as possible to real time and cost can be made, in case of a subsequent implementation. At the end of each project, the knowledge from Experience module is used to update the system knowledge base.

## CONSIDERATIONS CONCERNING PManager INTELLIGENT SYSTEM IMPLEMENTATION IN THE MANAGEMENT OF SOFTWARE DEVELOPMENT PLANNERS

A software development plan is realized, as a rule, by team work, the program product being conceived to be used by other persons than those who have developed it. Thus, the development of a user interface and the supply of the adherent documentation is required.

Moreover, the program product shall be properly tested on various platforms (hardware structures and operating systems) before delivery.

At present, software development projects imply high costs.

As witnessed by experience, in most cases, the software modules designed by the members of the team at software developing companies are supplied with delay, and are not reliable. This generally happens on account of the fact that the problem which the program product has to solve is not correctly defined, which finally leads either to the significant exceeding of the assigned budget, or to the cancelling of some software projects.

Software non-reliability has other causes than the non-reliability of the mechanical or electrical systems, which are due to the physical characteristics modifications in time. Errors in such cases appear due to design or coding errors. A program product may properly operate for some time because the parts containing errors are not used. Analyzing the present time situation of the evolution of software development projects, the following requirements shall be taken into account:

• The drawing up of an efficient planning of the tasks, according to the constraint: "As Soon As Possible";
• The shortening of the life cycle of software development planners;
• The development of an efficient mechanism for the follow-up, reporting, and control of the evolution of software development planners.

Thus, PManager Intelligent System proves useful for the creation of some actual sample planners, being provided with information supplied by unexpected situations in the knowledge base.

By highlighting the main stages that have been passed through for software products manufacture and maintenance, the life cycle is a simplified graphical representation, enabling the structuring of the tasks required by a software development plan.

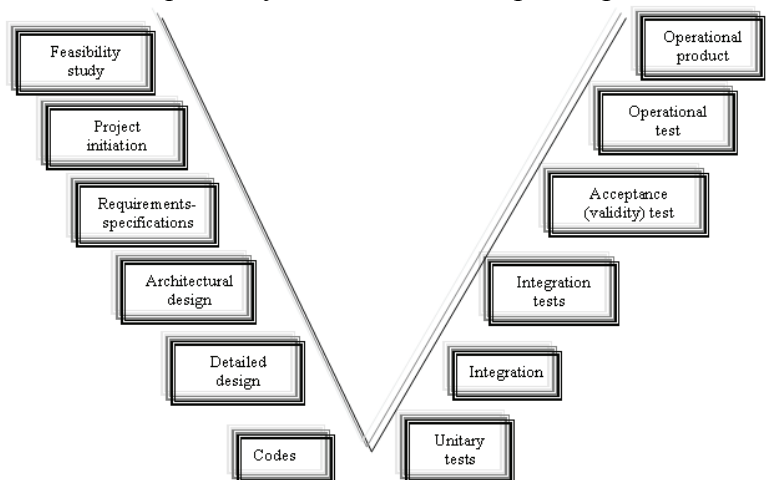

Figure 1: "V" life cycle

As far as the case study presented in this paper is concerned, there has been taken into consideration a certain type of software development project conceived according to "V" life cycle. Figure 1, "V" life cycle represents a traditional, well defined model, with sequential stages. The passage to a new phase is made only after the complete closing of the next phase. Within "V" life cycle (well known in technical literature) only one delivery is made.

Below there follows the presentation of PManager Intelligent System implementation, by means of the 12 steps of the usage own algorithm follow up [7][8]:

1. The tasks planner of the software development project is initially created in Microsoft Project according to the probable periods obtained from previous experience.

2. From Microsoft Project there can be obtained PERT (PERT Chart or Network Diagram) diagram on the bases of which CPM standard network is achieved, out of which the source and destination knots for each task will be obtained – data required for PManager Intelligent System implementation.

3. At the same time with the saved baseline planner, the "Earned Value" method is automatically activated, and an estimation of the manufacture cost as function of the specific situations can be obtained.

4. The information has been further imported in the fields of PManager Intelligent System, having the following significance: task code, source knot, destination knot, optimistic, probable, and pessimistic periods, respectively, and BCWS cost. PManager Intelligent System calculates the critical path on the optimistic, probable, pessimistic, and PERT sequences, respectively. There is also calculated the time buffer out of which transfers will be made for the optimization of some tasks with troubles, (Figure 2)

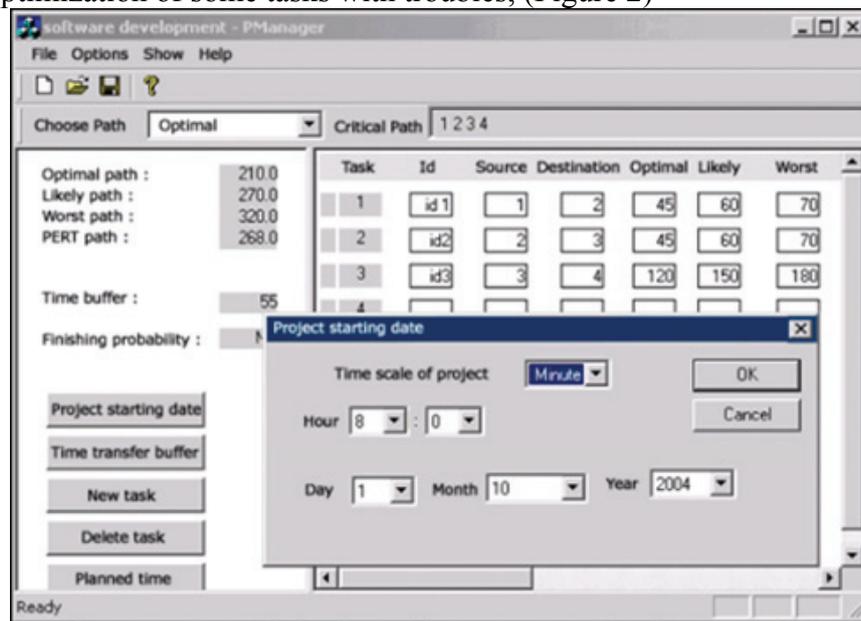

Figure 2:  PManager – the graphical interface

5. Schedule synchronization setting is made in PManager by entering the date of the project beginning, the project time unit, respectively, the planned (estimated) time for the project achievement, and the work program.

The time scheduled (estimated) for the project completion is required for calculating the project completion probability according to "Z" factor of PERT method. PManager updates the completion probability of the planner in real time after each adjustment of the tasks period, the user being permanently informed about the chances of time agreement. Moreover, if the completion period estimation deviates from the real value, being either too small, or too great, PManager Intelligent System warns the user by the following messages: "Too much waste of time", and  "Great risk of non-completion in due time".

6. Then, the system indicates, by means of a visual sign ">", the tasks that are in progress, monitoring the evolution of the planner according to the optimistic periods, and warns by a sound signal when getting closer to the end of the task, and by means of "!" visual sign it indicates the end of the task.

7. In case the user decides that a certain task does not range within the strictly operational period, a buffer transfer can be generated by pressing "Buffer time transfer" button, and as concerns the development of the knowledge base, the reason that has generated this transfer will be introduced.

Thus, as concerns task no. 6 from PManager planner (having the identification position from Gantt diagram from Microsoft Project id22, namely Know What to do indication, there has been made a buffer transfer of 30 minutes due to the wrong understanding of customer requirements.

Before time transfer achievement, the buffer was updated for 1917 value.

Following the 30 minutes transfer for id22 task, the buffer updated, having the new value 1905, as shown in the Table 1.

Table 1: Time buffer before/after the transfer in *PManager*

|  | Before | After |
|---|---|---|
| Time buffer (min) | 1917 | 1905 |

There can be observed that there did not take place a decrement of the time buffer of 30 minutes, it being updated on the basis of the same buffer calculation rule, following the schedule updating with the new data.

The tasks given below are examples that required further transfer:

• id31 (18) " SDD Remake", - 60 minutes buffer transfer due to "the remake of the interface between classes",

• id37 (28) "Code updating", - 60 minutes buffer transfer due to "the synchronization errors",

• id39 (32) "Code remake", - 60 minutes buffer due to "reorganization in classes".

8. Along with the planner evolution follow up by means of PManager Intelligent System, there takes place the adherent tracking within Microsoft Project. The activities that have been carried out according to the strictly operational (optimistic) period are updated in Microsoft Project according to these shorter periods. For instance, as concerns id-21 "Documents study" task, a completion period of 300 minutes has been obtained, according to the strictly operational period, the probable period from Microsoft Project planner being of 480 minutes . Most of the tasks from the planner could be forced to develop according to the strictly operational period, a drastically shortening of the planner being thus obtained, as has been anticipated. The real periods of the planner evolution have been subsequently updated in Microsoft Project, the costs evolution follow up being also achieved according to the Earned Value method. After a partial evaluation of the planner evolution, it has been found out that the daily „planning & tracking" task could be restricted to 45 minutes, the final cost results for this being as follows:

• For 13 sequences of „planning & tracking" (id3 task),

• BCWS total (the budgeted cost for the work scheduled) =1,950,000 ROL

• Following the shortening of the planner period there have resulted 11 „planning & tracking" sequences having BCWS total = 1,650,000 ROL and

• AC total (the actual cost for the planned work) = 1,237,500 ROL

• The total reduction is of  712,500 ROL.

There have been also updated the tasks periods to which buffer time transfer was made. Thus, id20 task – "Know What to Do" was planned during a strictly operational period of 45 minutes, but due to "the wrong understanding of customer requirements ", the task lasted for 75 minutes (30 minutes buffer transfer).

9. At the end of the planner tasks, the cost data will be collected according to the Earned Value method. These data will be imported to PManager Intelligent System, out of which the graphical evolution of costs minimization due to planner shortening will be obtained.

10. PManager Intelligent System ensures a diagnostic analysis of each task at the user's request by pressing the left side mouse button on the name of the task to be diagnosed, (Figure 3).

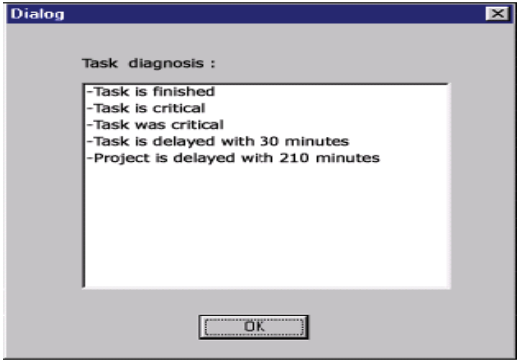

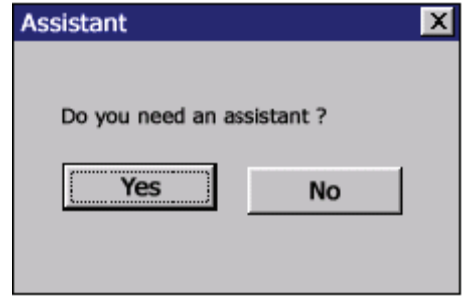

Figure 3: Diagnostic box for an task in PManager

Figure 4: PManager – Dialog box for calling the assistant

The diagnostic of the evolution of an task is achieved on the basis of the diagnostic tree. PManager updates the project network in real time, the planner being able to modify its configuration, the Critical Path, respectively. All these aspects are identified in the diagnostic box.

11. Another characteristic of PManager Intelligent System consists in its possibility to resort to an assistant which communicates interactively with the user in the course of the project progress, it always being a guide in supervising the planner, by forwarding ideas and suppositions with reference to the project tasks. (Figure 4)

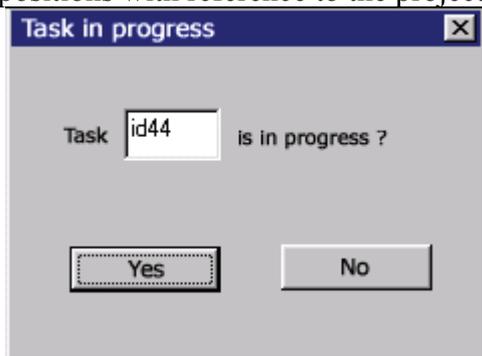

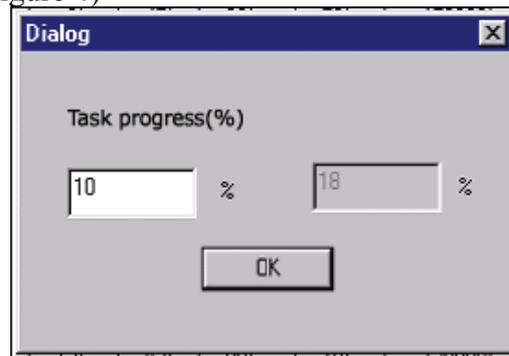

Figure 5: PManager – Dialog box that confirms the evolution

Figure 6: Dialog box for displaying the percentage of the task achievement

In case an assistant help is selected, it will send some warning messages concerning planner tasks status. The following situations have been considered: Figure 5 - Figure 9.

In "Task in progress" dialog box, the user confirms whether the task has been started or not. In case it has, the assistant generates another dialog box to see the percentage already achieved of that task.

In case of the dialog box in Figure 4, PManager System identifies by means of the synchronization procedure, the tasks which should develop, according to the initial planner. Thus, the system warns the user concerning the situation, asking for a confirmation. Figure 5

The right side cell (Figure 6) of this dialog box displays the percentage of the respective task achievement, in case the task developed according to the initial planner. In case a disagreement is found out, the real percentage is entered in the left planner. In case a disagreement is found out, the real percentage is entered in the left side cell.

In case an achievement percentage smaller than the planned one has been entered in the previous dialog box, the program suggests a time transfer from the buffer whose value is calculated so as to compensate the deviation of the task from the initial planner (Figure 7)

In case the time transfer from the buffer is accepted, the assistant generates the automatic opening of "Buffer time transfer" dialog box, where the user can validate or change the transfer suggested by the program. The user can also enter the reasons that generated the delay of the task (Figure 8).

Following the close of supervising and updating procedures, the assistant suggests a project analysis (Figure 9). The information from "Project analysis" dialog box is generated following the diagnosis of the evolution of all the tasks (startup, finished, non-startup).

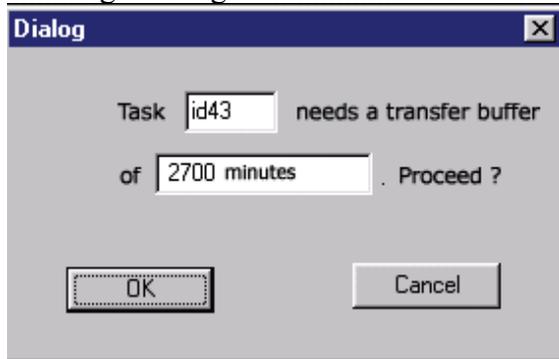

Figure 7: Dialog box for compensating the deviations

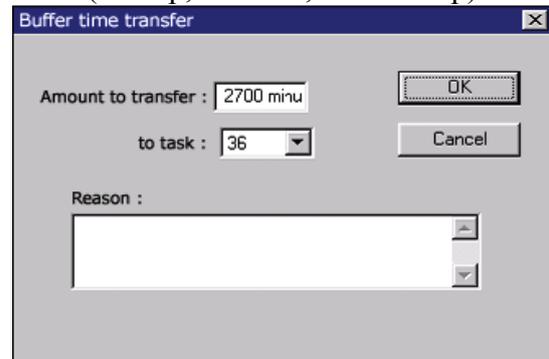

Figure 8: Dialog box for buffer transfer achievement

12. At a subsequent use of the planner, the knowledge base updated by means of "Experience" module can be selected (Figure 10).

Following knowledge base selection, the planner, which will have the same structure but different characteristics, can be entered. When entering the attributes of an task which required buffer time transfer, according to the knowledge base, PManager Intelligent System displays a warning box informing that the task required a transfer from the time reserve, the initial planning period, the period cumulated by transfer, and the reasons that generated this adjustment being specified.

The example refers to id22 task, "Know What to do", which required 30 minutes buffer transfer (from 45 minutes to 75 minutes, due to the wrong understanding of customer requirements), according to the model created by in the knowledge base, (Figure 11).

The knowledge cumulated by PManager Intelligent System during the evolution of a planner is stored as frames.

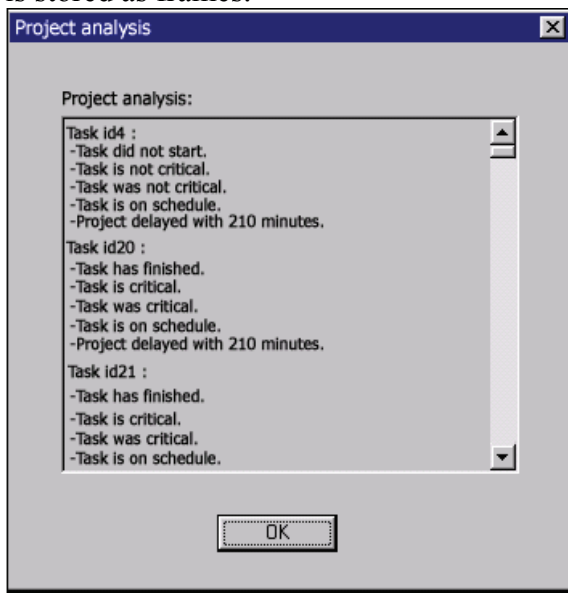

Figure 9: PManager - The analysis of the project evolution at a certain moment

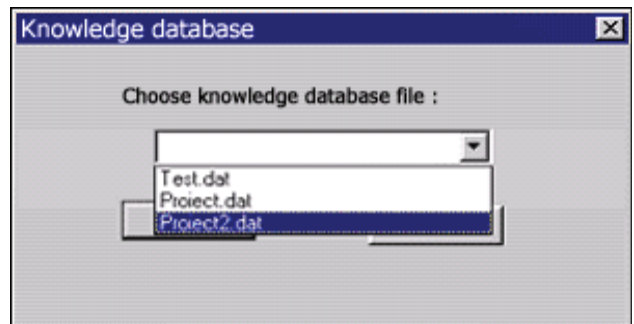

Figure 10: PManager Knowledge base selection

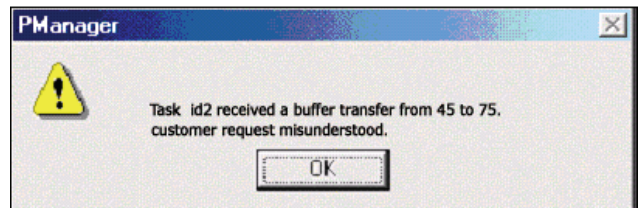

Figure 11:PManager – Warning referring to the change of the period of an task within the previous evolution of a planner

This file memorizes all the tasks which could not develop according to the optimistic period, the frame of each task supplying the following information:
• the name (identifier) of the task to which a time transfer has been made
• the value of the optimistic time (initially planned)
• the time transfer which has been applied to the task
• the reasons which have generated the extension of the task
The user can also visualize the knowledge base for carrying out a diagnostic analysis of the evolution of Critical Path sequence. The frame of this file has the following structure:
• critical path sequence
• the task to which a buffer transfer has been made and which generated the change of the critical path
• the value of the time transfer with which the task was shifted
• critical path period

## CONCLUSION

The real time follow up, synchronization, and updating of planners evolution by means of software tools, represents one of the major concerns in the field of the scheduling problems. PManager Intelligent System has been specially created for the purpose, additionally offering the possibility of centralized projects (operations) scheduling and follow up.

PManager Intelligent System is provided with a user-friendly interface, which offers suggestions and changes in real time for all the unexpected situations that might happen as a result of the use of all the methods created during the phase of architectural design.

Following PManager Intelligent System implementation for a series of 10 projects of the same type, at the same company for software projects development, there has been obtained an efficient management of these projects life cycle as specified below:

1. by means of PManager System there has been achieved the shortening of the completion period and the follow up of all the projects during their evolution, real time adjustments being offered for each deviation from the initial planners.

2. there has been created a knowledge base useful within the Intelligent System, which comes up against the situations that cannot be foreseen as early as the initial planning and estimations.

3. based on the created model, the company where the Intelligent System has been implemented has improved its essential parameters of software Projects Management, namely:
• It develops projects that comply with customer requirements.
• It delivers the final product within a shorter term.
• The production costs have been reduced.

4. the resulting cost ratios came up against the economic-financial analysis of the company.